\input ./style/arxiv-vmsta.cfg
\documentclass[numbers,compress,v1.0.1]{vmsta}
\usepackage{vmkcol}

\volume{4}% Updated by VTEXPTS2LaTeX.exe, 30.06.2017 13:01
\issue{2}% Updated by VTEXPTS2LaTeX.exe, 30.06.2017 13:01
\pubyear{2017}
\firstpage{109}% Updated by VTEXPTS2LaTeX.exe, 30.06.2017 13:01
\lastpage{125}% Updated by VTEXPTS2LaTeX.exe, 30.06.2017 13:01
\doi{10.15559/17-VMSTA77}% Updated by VTEXPTS2LaTeX.exe, 21.04.2017
\startlocaldefs

\allowdisplaybreaks

\newtheorem{theorem}{Theorem}
\newtheorem{proposition}{Proposition}
\theoremstyle{definition}
\newtheorem{definition}{Definition}

\newcommand{\ncm}{\newcommand}
\ncm{\bfm}[1]{\mbox{\boldmath$#1$}}
\ncm{\scbfm}[1]{\mbox{\scriptsize\boldmath$#1$}}
\ncm{\sbfm}[1]{\mbox{\footnotesize\boldmath$#1$}}
\ncm{\scr}[1]{\mbox{\footnotesize #1}}
\ncm{\tin}[1]{\mbox{\footnotesize #1}}
\ncm{\bfmscr}[1]{\mbox{\scriptsize{\boldmath$ #1$}}}

\ncm{\R}{{\mathbb{R}}}
\ncm{\Z}{{\mathbb{Z}}}
\ncm{\T}{{\mathbb{T}}}
\ncm{\Smath}{{\mathbb{S}}}
\ncm{\N}{{\mathbb{N}}}
\ncm{\A}{{\mathbb{A}}}
\ncm{\amath}{\bfm{a}}
\ncm{\V}{{\mathbb{V}}}
\ncm{\Hap}{{\mathbb{H}}}
\ncm{\MMD}{{\mathbb{MD}}}

\ncm{\cA}{{\cal A}}
\ncm{\cB}{{\cal B}}
\ncm{\cC}{{\cal C}}
\ncm{\calF}{{\cal F}}
\ncm{\cD}{{\cal D}}
\ncm{\cG}{{\cal G}}
\ncm{\cN}{{\cal N}}
\ncm{\cI}{{\cal I}}
\ncm{\cJ}{{\cal J}}
\ncm{\cH}{{\cal H}}
\ncm{\cV}{{\cal V}}
\ncm{\cW}{{\cal W}}
\ncm{\cT}{{\cal T}}
\ncm{\cX}{{\cal X}}
\ncm{\cQ}{{\cal Q}}
\ncm{\cR}{{\cal R}}
\ncm{\cS}{{\cal S}}
\ncm{\cM}{{\cal M}}
\ncm{\cU}{{\cal U}}
\ncm{\cP}{{\cal P}}
\ncm{\cO}{{\cal O}}
\ncm{\cPzer}{{\cal P}_0}
\ncm{\cPone}{{\cal P}_1}
\ncm{\cPk}{{\cP_{\mbox{\scr{known}}}}}
\ncm{\cF}{{\cal F}}
\ncm{\cE}{{\cal E}}
\ncm{\cMD}{{\cal MD}}
\ncm{\tcV}{\tilde{\cal V}}
\ncm{\cCobs}{{\cal C}_{\scr{obs}}}

\ncm{\Om}{\varOmega}
\ncm{\om}{\omega}
\ncm{\va}{\varepsilon}
\ncm{\vam}{\varepsilon_{\scr{max}}}
\ncm{\de}{\delta}
\ncm{\De}{\Delta}
\ncm{\ga}{\gamma}
\ncm{\Ga}{\varGamma}
\ncm{\la}{\lambda}
\ncm{\ka}{\kappa}
\ncm{\si}{\sigma}
\ncm{\Si}{\varSigma}
\ncm{\La}{\varLambda}
\ncm{\eps}{\epsilon}

\ncm{\bY}{\bfm{Y}}
\ncm{\bA}{\bfm{A}}
\ncm{\bB}{\bfm{B}}
\ncm{\bC}{\bfm{C}}
\ncm{\bD}{\bfm{D}}
\ncm{\bF}{\bfm{F}}
\ncm{\bI}{\bfm{I}}
\ncm{\bZ}{\bfm{Z}}
\ncm{\bG}{\bfm{G}}
\ncm{\bH}{\bfm{H}}
\ncm{\bL}{\bfm{L}}
\ncm{\bQ}{\bfm{Q}}
\ncm{\bS}{\bfm{S}}
\ncm{\bT}{\bfm{T}}
\ncm{\bU}{\bfm{U}}
\ncm{\bX}{\bfm{X}}
\ncm{\bM}{\bfm{M}}
\ncm{\bN}{\bfm{N}}
\ncm{\bu}{\bfm{u}}
\ncm{\bv}{\bfm{v}}
\ncm{\bw}{\bfm{w}}
\ncm{\bwpr}{\bfm{w}^\prime}
\ncm{\bhp}{\bfm{h}^\prime}
\ncm{\bd}{\bfm{d}}
\ncm{\bm}{\bfm{m}}
\ncm{\bh}{\bfm{h}}
\ncm{\bn}{\bfm{n}}
\ncm{\bb}{\bfm{b}}
\ncm{\bg}{\bfm{g}}
\ncm{\be}{\bfm{e}}
\ncm{\bl}{\bfm{l}}
\ncm{\bq}{\bfm{q}}
\ncm{\bs}{\bfm{s}}
\ncm{\bx}{\bfm{x}}
\ncm{\by}{\bfm{y}}
\ncm{\bz}{\bfm{z}}
\ncm{\balp}{\bfm{\alpha}}
\ncm{\bbe}{\bfm{\beta}}
\ncm{\bxi}{\bfm{\xi}}
\ncm{\bth}{\bfm{\theta}}
\ncm{\bom}{\bfm{\om}}
\ncm{\bmu}{\bfm{\mu}}
\ncm{\bde}{\bfm{\de}}
\ncm{\bva}{\bfm{\va}}
\ncm{\beps}{\bfm{\eps}}
\ncm{\bpi}{\bfm{\pi}}
\ncm{\brho}{\bfm{\rho}}
\ncm{\boldeta}{\bfm{\eta}}
\ncm{\bphi}{\bfm{\phi}}
\ncm{\bpsi}{\bfm{\psi}}
\ncm{\bLa}{\bfm{\varLambda}}
\ncm{\bPi}{\bfm{\varPi}}
\ncm{\bsi}{\bfm{\si}}
\ncm{\bka}{\bfm{\ka}}
\ncm{\bSi}{\bfm{\Si}}
\ncm{\bone}{\bfm{1}}
\ncm{\bzero}{\bfm{0}}

\ncm{\sbM}{\sbfm{M}}
\ncm{\sbe}{\sbfm{e}}
\ncm{\sbx}{\sbfm{x}}
\ncm{\sby}{\sbfm{y}}
\ncm{\sbu}{\sbfm{u}}
\ncm{\sbv}{\sbfm{v}}
\ncm{\sbw}{\sbfm{w}}
\ncm{\sbpsi}{\sbfm{\psi}}
\ncm{\sbka}{\sbfm{\kappa}}

\ncm{\heta}{\hat{\eta}}
\ncm{\hth}{\hat{\theta}}
\ncm{\hbth}{\hat{\bth}}
\ncm{\hatp}{\hat{p}}
\ncm{\ha}{\hat{a}}
\ncm{\hq}{\hat{q}}
\ncm{\hs}{\hat{s}}
\ncm{\hz}{\hat{z}}
\ncm{\hN}{\hat{N}}
\ncm{\hF}{\hat{F}}
\ncm{\hI}{\hat{I}}
\ncm{\hP}{\hat{P}}
\ncm{\hS}{\hat{S}}
\ncm{\hsi}{\hat{\si}}
\ncm{\hSi}{\hat{\Si}}
\ncm{\hpi}{\hat{\pi}}
\ncm{\hxi}{\hat{\xi}}
\ncm{\hpsi}{\hat{\psi}}
\ncm{\hbpsi}{\hat{\bpsi}}
\ncm{\hbka}{\hat{\bfm{\kappa}}}
\ncm{\hbxi}{\hat{\bxi}}
\ncm{\htau}{\hat{\tau}}
\ncm{\hbq}{\hat{\bq}}
\ncm{\hbD}{\hat{\bD}}
\ncm{\hbSi}{\hat{\bfm{\varSigma}}}

\ncm{\mast}{m^\ast}
\ncm{\cast}{c^\ast}
\ncm{\fast}{f^\ast}
\ncm{\siast}{\si^\ast}
\ncm{\psiast}{\psi^\ast}
\ncm{\tsiast}{\tilde{\si}^\ast}
\ncm{\alfast}{\alpha^\ast}
\ncm{\tkaast}{\tilde{\kappa}^\ast}

\ncm{\wpr}{w^\prime}
\ncm{\jp}{j^\prime}
\ncm{\ip}{i^\prime}
\ncm{\Ip}{I^\prime}
\ncm{\kp}{k^\prime}
\ncm{\hp}{h^\prime}
\ncm{\lp}{l^\prime}
\ncm{\np}{n^\prime}
\ncm{\Gp}{G^\prime}
\ncm{\Mp}{M^\prime}
\ncm{\Cp}{C^\prime}
\ncm{\ap}{a^\prime}
\ncm{\vp}{v^\prime}
\ncm{\up}{u^\prime}
\ncm{\npr}{n^\prime}
\ncm{\Npr}{N^\prime}
\ncm{\xp}{x^\prime}
\ncm{\yp}{y^\prime}
\ncm{\zp}{z^\prime}
\ncm{\phpr}{\phi^\prime}

\ncm{\wbis}{w^{\prime\prime}}

\ncm{\tih}{\tilde{h}}
\ncm{\tZ}{\tilde{Z}}
\ncm{\tA}{\tilde{A}}
\ncm{\tF}{\tilde{F}}
\ncm{\tI}{\tilde{I}}
\ncm{\tmu}{\tilde{\mu}}
\ncm{\tOm}{\tilde{\varOmega}}
\ncm{\tnu}{\tilde{\nu}}
\ncm{\tsi}{\tilde{\sigma}}
\ncm{\tal}{\tilde{\alpha}}
\ncm{\tbeta}{\tilde{\beta}}
\ncm{\tde}{\tilde{\delta}}
\ncm{\txi}{\tilde{\xi}}
\ncm{\tmathV}{\tilde{\V}}
\ncm{\tV}{\tilde{V}}
\ncm{\tu}{\tilde{u}}
\ncm{\tw}{\tilde{w}}
\ncm{\twpr}{\tilde{w}^\prime}
\ncm{\td}{\tilde{d}}
\ncm{\tp}{\tilde{p}}
\ncm{\tf}{\tilde{f}}
\ncm{\tn}{\tilde{n}}
\ncm{\tS}{\tilde{S}}
\ncm{\tL}{\tilde{L}}
\ncm{\tl}{\tilde{l}}
\ncm{\tP}{\tilde{P}}
\ncm{\tSmath}{\tilde{\mathbb{S}}}
\ncm{\tT}{\tilde{T}}
\ncm{\tK}{\tilde{K}}
\ncm{\tka}{\tilde{\ka}}
\ncm{\tva}{\tilde{\va}}
\ncm{\tla}{\tilde{\la}}
\ncm{\tpi}{\tilde{\pi}}
\ncm{\tbe}{\tilde{\bfm{e}}}
\ncm{\tbom}{\tilde{\bfm{\om}}}
\ncm{\tbxi}{\tilde{\bfm{\xi}}}
\ncm{\tbg}{\tilde{\bg}}
\ncm{\tbb}{\tilde{\bb}}

\ncm{\baW}{\bar{W}}
\ncm{\bacW}{\bar{\cW}}
\ncm{\bacV}{\bar{\cV}}
\ncm{\ban}{\bar{n}}
\ncm{\bap}{\bar{p}}
\ncm{\bav}{\bar{v}}
\ncm{\baw}{\bar{w}}
\ncm{\baZ}{\bar{Z}}
\ncm{\baY}{\bar{Y}}
\ncm{\baS}{\bar{S}}
\ncm{\baH}{\bar{H}}
\ncm{\baD}{\bar{D}}
\ncm{\baC}{\bar{C}}
\ncm{\baJ}{\bar{J}}
\ncm{\baO}{\bar{O}}
\ncm{\bah}{\bar{h}}
\ncm{\bal}{\bar{l}}
\ncm{\bam}{\bar{m}}
\ncm{\bae}{\bar{e}}
\ncm{\bacR}{\bar{{\cal R}}}
\ncm{\bacP}{\bar{{\cal P}}}
\ncm{\baka}{\bar{\kappa}}
\ncm{\bamu}{\bar{\mu}}
\ncm{\banu}{\bar{\nu}}
\ncm{\bala}{\bar{\la}}
\ncm{\baga}{\bar{\ga}}
\ncm{\bath}{\bar{\theta}}
\ncm{\baOR}{\overline{\mbox{OR}}}
\ncm{\baRR}{\overline{\mbox{RR}}}
\ncm{\babpsi}{\bar{\bpsi}}
\ncm{\babD}{\bar{\bD}}
\ncm{\babSi}{\bar{\bSi}}

\ncm{\chnu}{\check{\nu}}

\ncm{\Leq}{\, \stackrel{\cal L} =}
\ncm{\Lto}{\, \stackrel{\cal L} \longrightarrow}
\ncm{\pto}{\, \stackrel{p} \longrightarrow}
\ncm{\asto}{\, \stackrel{\rm a.s.} \longrightarrow}
\ncm{\Cov}{\mbox{Cov}}
\ncm{\Var}{\mbox{Var}}
\ncm{\sameord}{\stackrel{\cup}{{\scriptstyle\cap}}}

\ncm{\ith}{i^{\scr{th}}}
\ncm{\jth}{j^{\scr{th}}}
\ncm{\kth}{k^{\scr{th}}}
\ncm{\lth}{l^{\scr{th}}}
\ncm{\Bin}{\mbox{Bin}}
\ncm{\Hyp}{\mbox{Hyp}}
\ncm{\OR}{\mbox{OR}}
\ncm{\sOR}{\scr{OR}}
\ncm{\RR}{\mbox{RR}}
\ncm{\AP}{\mbox{AP}}
\ncm{\PAP}{\mbox{PAP}}
\ncm{\PAF}{\mbox{PAF}}
\ncm{\SE}{\mbox{SE}}
\ncm{\CI}{\mbox{CI}}
\ncm{\SI}{\mbox{SI}}
\ncm{\ORI}{\mbox{ORI}}
\ncm{\EOR}{\mbox{EOR}}
\ncm{\UEOR}{\mbox{UnadjEOR}}
\ncm{\sUEOR}{\scr{UnadjEOR}}
\ncm{\hAP}{\widehat{\mbox{AP}}}
\ncm{\hAPast}{\widehat{\mbox{AP}}^{\ast}}
\ncm{\hxiast}{\hxi^{\ast}}
\ncm{\hbpsiast}{\widehat{\bpsi}^\ast}
\ncm{\RRt}{\mbox{\tiny RR}}
\ncm{\RRf}{\mbox{\footnotesize RR}}
\ncm{\ORf}{\mbox{\footnotesize OR}}

\ncm{\beq}{
\begin{equation}
} \ncm{\eeq} {
\end{equation}
}
\ncm{\beqr}{
\begin{eqnarray}
} \ncm{\eeqr} {
\end{eqnarray}
}
\ncm{\beqrn}{
\begin{eqnarray*}
} \ncm{\eeqrn} {
\end{eqnarray*}
}
\ncm\rthm[1]{\ref{#1}}
\ncm\lb[1]{\label{#1}}
\ncm\re[1]{(\ref{#1})}
\ncm{\slut}{
{\unskip\nobreak\hfill\penalty100\hskip1em\vadjust{}\nobreak
\hfill\mbox{$\Box$}\parfillskip=0pt\finalhyphendemerits=0}}
\endlocaldefs

\begin{document}
\begin{frontmatter}

\title{Quantifying and estimating additive measures of interaction from
case-control data}
%\author[]{\inits{}\fnm{}\snm{}\corref{cor1}}\email{}
%\cortext[cor1]{Corresponding author.}

%\author[]{\inits{}\fnm{}\snm{}}\email{}

%\fnref{f1}
%\fntext[]{Some remarks}

%\address[]{}
%\address[]{}

%\markboth{A. Authors}{Title}

\author[a]{\inits{O.}\fnm{Ola}\snm{H{\"o}ssjer}\corref{cor1}}\email
{ola@math.su.se}
\cortext[cor1]{Corresponding author.}
\author[b]{\inits{L.}\fnm{Lars}\snm{Alfredsson}}\email{Lars.Alfredsson@ki.se}
\author[b]{\inits{A.}\fnm{Anna Karin}\snm{Hedstr{\"o}m}}\email
{anna.hedstrom@ki.se}
\author[c]{\inits{M.}\fnm{Magnus}\snm{Lekman}}\email{magnus.lekman@gmail.com}
\author[c]{\inits{I.}\fnm{Ingrid}\snm{Kockum}}\email{Ingrid.Kockum@ki.se}
\author[c]{\inits{T.}\fnm{Tomas}\snm{Olsson}}\email{Tomas.Olsson@ki.se}
\address[a]{Department of Mathematics, Stockholm University, Stockholm, Sweden}
\address[b]{Institute of Environmental Medicine, Karolinska Institutet,
Stockholm, Sweden}
\address[c]{Department of Clinical Neuroscience, Karolinska Institutet,
Stockholm, Sweden}

\markboth{O.\ H{\"o}ssjer et al.}{Quantifying and estimating additive measures of interaction from
case-control data}

%\begin{abstract}
%\end{abstract}

%\begin{keywords}
%\kwd{}
%\kwd{}
%\kwd{}
%\kwd{}
%\end{keywords}
%\begin{keywords}[2010]% [PACS], [JEL]
%\kwd{}
%\kwd{}
%\kwd{}
%\kwd{}
%\end{keywords}

\begin{abstract}
In this paper we develop a general framework for quantifying how binary
risk factors jointly influence a binary outcome. Our key result is an
additive expansion of odds ratios as a sum of marginal effects and
interaction terms of varying order. These odds ratio expansions are
used for estimating the excess odds ratio, attributable proportion and
synergy index for a case-control dataset by means of maximum likelihood
from a logistic regression model. The confidence intervals associated
with these estimates of joint effects and interaction of risk factors
rely on the delta method. Our methodology is illustrated with a large
Nordic meta dataset for multiple sclerosis. It combines four studies,
with a total of 6265 cases and 8401 controls. It has three risk factors
(smoking and two genetic factors) and a number of other confounding variables.
\end{abstract}


\received{22 March 2017}% Updated by VTEXPTS2LaTeX.exe, 21.04.2017 11:18
\revised{12 April 2017}% Updated by VTEXPTS2LaTeX.exe, 21.04.2017 11:18
\accepted{12 April 2017}% Updated by VTEXPTS2LaTeX.exe, 21.04.2017 11:18
\publishedonline{26 April 2017}
\end{frontmatter}

\section{Introduction}\vspace*{-1.2pt}

Many complex diseases are influenced by a number of risk factors that
interact in a complicated way. This is often quantified by means of a
regression model with affection status of a given disease as binary
response, whereas the risk factors and possibly some other variables
are chosen as covariates. Logistic regression models have often been
used to quantify main effects and strength of interaction among the
risk factors with regards to disease. There are several reasons for
this. The logistic transformation is first of all the canonical link of
a generalized linear model with a binomially distributed response, and
the parameters of this model have a straightforward multiplicative odds
interpretation \cite{McC89}. A second reason is that
many epidemiological datasets are collected retrospectively based on
outcomes rather than on covariates. In particular, it is well known
that many parameters of the logistic regression model can be estimated
consistently for case-control studies under suitable sampling
assumptions on the cases and controls \cite{Pre79}. There
are also additive models of joint effects and interaction. They have
recently gained in popularity, since they are believed to approximate a
biological system with causal mechanisms more accurately than
multiplicative odds \cite{Van08,Rot12}. The
additive measures of interaction are functions of relative risks
between individuals with or without exposure to the risk factors, such
as the relative excess of risk due to interaction (RERI), the
attributable proportion (AP) due to interaction or the synergy index
(SI) \cite{Gre93,Skr03}.
Although these relative risks cannot be estimated consistently for
case-control data, they are well approximated by estimable odds ratios
when the disease risk is small \cite{Hos92,Ass96,And15}. When relative risks are replaced by
odds ratios in the expressions for RERI, AP and SI, they correspond to
additive odds models of interaction.

In previous work \cite{Hos17} we developed a unified
theory for quantifying and estimating different measures of marginal
effects, joint effects and interaction among risk factors on a
multiplicative, additive, additive odds or some other scale. We also
described how to estimate and produce confidence intervals for these
quantities from a prospective study, with data sampled based on their
covariates, or from a case-control study. Traditional definitions of
attributable proportion have an unbounded negative range \cite{Kno11,Van13,Lek14}.
In \cite{Hos17}, we introduced a novel normalization of AP that guarantees a
range between $-1$ and 1, with negative or positive values depending on
whether there is synergism or antagonism between the risk factors.

In this paper we concentrate on case-control data and additive odds
measures of main effects, joint effects and interaction. Our main
result is to express the odds ratio of the risk factors as a sum of
terms, which include their main effects and different orders of
interaction, when the effect of other confounding covariates is
controlled for. In this way we extend and unify some previously used
measures of interaction \cite{Pet13,Kat17,Hos17}. In order to find confidence intervals (CIs)
for the attributable proportion, synergy index and excess odds ratio
due to joint effects and interaction from case-control data, we first
estimate the parameters of a logistic regression model by maximum
likelihood. Then we use the delta method \cite{Rot80,Cas02} for an appropriate transformation of these measures
of joint effects and interaction in order to find their standard errors
and CIs on the new scale, before transforming back to the original odds scale.

The paper is organized as follows: In Section~\ref{Sec:Model} we define
a logistic regression model that includes marginal and interaction
effects for the risk factors of interest,\vadjust{\eject} as well as marginal effects
of other confounding covariates. The additive odds measures of joint
effects and interaction are defined in Section~\ref{Sec:xi},
and the procedures for estimating them are given in Section~\ref{Sec:Inf}.
Our methodology is illustrated in Section~\ref{Sec:Data} for
a multiple sclerosis (MS) dataset from Hedstr\"{o}m et al. \cite{Hed17}, and
a concluding discussion appears in Section~\ref{Sec:Disc}.

\section{A logistic regression model}\lb{Sec:Model}

Let $Y\in\{0,1\}$ be a binary outcome variable, with $Y=1$ for
individuals that carry a certain disease, and $Y=0$ for those that do
not. Consider a large population, and let
\beq
\theta_{\sbfm{x}} = P(Y=1|\bx)
\lb{thx}
\eeq
be the disease probability of a randomly chosen individual. It is
assumed to be a function of $p+q$ covariates $\bx=(\bv,\bz)=(v_1,\ldots
,v_p,z_1,\ldots,z_q)$,
of which the first $p$ are binary risk factors $v_1,\ldots,v_p$ of main
interest, with $v_j=1$ indicating presence and $v_j=0$ absence of each
such factor $j$.
The other $q$ covariates $z_1,\ldots,z_q$ are not necessarily binary,
and they are included in the model as possible confounders. We will
parametrize the disease probability on a logit scale
\beq
\mbox{logit}\, \theta_{\sbfm{x}} = \log\frac{\theta_{\sbfm{x}}}{1-\theta
_{\sbfm{x}}}
= \kappa_0 + \sum_{\sbfm{0}<\sbfm{w}\le\sbv} \psi_{\sbfm{w}} + \sum_{j=1}^q \ka_j z_j,
\lb{etax}
\eeq
with $\bw=(w_1,\ldots,w_p)$ a binary vector and $\bzero= (0,\ldots,0)$
a vector with zero components.
Inequalities between vectors are interpreted componentwise, so that the
first sum in \re{etax} is taken over all nonzero vectors $\bw$ such
that $w_j\le v_j$,
with $w_j=1$ for at least one factor $j$. This sum equals 0 when $\bv
=\bzero$.

The logistic model in \re{etax} is saturated for the binary risk
factors, including all orders of their interaction, whereas it is linear
in the other covariates \cite{Agr92,Agr13}.
In particular, if $v_j=1$ and all other components of $\bv$ are zero,
$\psi_{\sbfm{v}}$ quantifies the marginal effect of factor $j$ in
absence of the others, whereas $\psi_{\sbfm{v}}$ for $\bv$ such that
$|\bv|=\sum_j v_j\ge2$ quantifies interaction of order $|\bv|$ among
those factors $j$ for which $v_j=1$.

It is assumed that $\bka=(\ka_0,\ka_1,\ldots,\ka_q)$ are nuisance
parameters, whereas
\beq
\bpsi= (\bpsi_{\sbfm{v}}; \bzero< \bv\le\bone)
\lb{bpsi}
\eeq
are the $2^p-1$ structural parameters of main interest that quantify
marginal effects and interaction among the $p$ risk factors, with
$\bone= (1,\ldots,1)$.
Since they are defined on a logit scale, it is possible to estimate
them from suitable case-control datasets. In the next section we will
develop alternative measures of marginal effects, joint effects and
interaction that are functions of $\bpsi$, but expressed in terms of
odds ratios.

\section{Measures of joint effects and interaction}\lb{Sec:xi}

Our purpose is to estimate and produce confidence intervals for parameters
\beq
\xi= \xi(\bpsi)
\lb{xi}
\eeq
that quantify marginal effects, joint effects or interaction between a
subset $J\subset\{1,\ldots,p\}$
of factors, when the levels of the remaining factors in $K = \{1,\ldots
,p\}\setminus J$,\vadjust{\eject} and the $q$ other confounder variables are controlled
for. It is assumed that
\re{xi} only involves the structural parameters $\bpsi$, not the
nuisance parameters $\bka$.
The reason is that we consider parameters \re{xi} that are functions of
odds ratios
\beq
\OR_{\sbfm{v}} = \frac{\theta_{(\sbfm{v},\sbfm{z})}/(1-\theta_{(\sbfm
{v},\sbfm{z})})}{\theta_{(\sbfm{0},\sbfm{z})}/(1-\theta_{(\sbfm{0},\sbfm{z})})}
= \exp \biggl(\sum_{\sbfm{0}<\sbfm{w}\le\sbv} \psi_{\sbfm{w}} \biggr),
\lb{OR}
\eeq
i.e.\ ratios between the disease odds of two subjects with identical
confounding variables $\bz$ but different risk exposures $\bv$ and
$\bzero$ respectively,
for some or all of the $2^p-1$ vectors $\bzero< \bv\le\bone$. In
particular, when the disease risks in \re{thx} are small, the odds
ratio in \re{OR} approximates the relative risk
\beq
\RR_{(\sbfm{v},\sbfm{z})} = \frac{\theta_{(\sbfm{v},\sbfm{z})}}{\theta
_{(\sbfm{0},\sbfm{z})}}
\lb{RR}
\eeq
well.

After possible reordering of factors we may assume without loss of
generality that those in $J$ come first, so that $\bv=(\bv_J,\bv_K)$,
with $\bv_J = (v_j;\, j\in J)$ and $\bv_K = (v_j;\, j\in K)$
being the exposure levels of the factors in $J$ and $K$ respectively.
We will consider linear combinations
of the odds ratios \re{OR}, and the following concept will be central
for the rest of the paper:

\begin{definition}
Suppose the exposure levels $\bv_K$ of the confounding risk factors are
fixed. For any $\bv_J$ we introduce the additive increment
\beq
\varDelta^{(\sbv_J,\sbv_K)} \OR= \sum_{\sbfm{0}\le\sbw\le\sbv_J}
(-1)^{|\sbv_J-\sbw|}\OR_{(\sbw,\sbv_K)}
\lb{DeltasbvJ}
\eeq
of the odds ratio of order $\bv_J$, where summation on the right-hand
side of \re{DeltasbvJ} is over all binary vectors $\bw=(w_j;\, j\in J)$
of length $|J|$ whose coordinates do not exceed those of $\bv_J$.
\end{definition}

When $|\bv_J|=0$, \re{DeltasbvJ} is the odds ratio when none of the
risk variables in $J$ are turned on, and when $|\bv_J|=1$ it is the
marginal odds ratio increment of one single factor. For $|\bv_J|\ge2$
we interpret \re{DeltasbvJ} as a measure of additive odds interaction
among those factors $\{j\in J;\, v_j=1\}$ in $J$ that are at risk, when
all interaction terms of lower order among these factors have been
removed. More explicitly,
the terms of order 0, 1, 2 and 3 in \re{DeltasbvJ} have the form
\beq
\varDelta^{(\sbv_J,\sbv_K)} \OR\,{=}
\begin{cases}
\OR_{(\sbfm{0},\sbv_K)}, & \bv_J = \bzero,\\
\OR_{(\sbfm{e}_j,\sbv_K)} - \OR_{(\sbfm{0},\sbv_{K})}, & \bv_J=\bfm
{e}_j,\\
\OR_{(\sbfm{e}_{jk},\sbv_K)} - \OR_{(\sbe_j,\sbv_{K})} - \OR_{(\sbe
_k,\sbv_{K})} + \OR_{(\sbfm{0},\sbv_{K})}, & \bv_J=\bfm{e}_{jk},\\
\OR_{(\sbfm{e}_{jkl},\sbv_K)} - \OR_{(\sbe_{jk},\sbv_{K})} - \OR_{(\sbe
_{jl},\sbv_{K})} & \\
\quad -\, \OR_{(\sbe_{kl},\sbv_{K})} + \OR_{(\sbe_j,\sbv_{K})} + \OR_{(\sbe
_k,\sbv_{K})} & \\
\quad +\, \OR_{(\sbe_l,\sbv_{K})} - \OR_{(\sbfm{0},\sbv_{K})}, & \bv_J=\bfm
{e}_{jkl},
\end{cases}
\lb{DeltasbvJEx}
\eeq
where $\bfm{e}_j$, $\bfm{e}_{jk}$ and $\bfm{e}_{jkl}$ are binary
vectors of length $|J|$ with zero components, except in positions $\{j\}
$, $\{j,k\}$ and $\{j,k,l\}$ respectively.

It is possible to give another interpretation of \re{DeltasbvJ}. Let
$\bw= (w_j; \, j\in J)$ be a vector with $0\le w_j\le1$,
and extend the domain of the first $|J|$ components of the odds ratio
$\OR_{(\sbw,\sbv_K)}$ to $[0,1]^{|J|}$.
If $\bv_J$ is a binary vector of length $|\bv_J|=l>0$ with nonzero
components $v_{j_1}=\cdots= v_{j_l}=1$,
then $\varDelta^{(\sbv_J,\sbv_K)} \OR$ is a finite difference
approximation of $\partial^{l} \OR_{(\sbw,\sbv_K)}/(\partial
w_{j_1}\ldots\partial w_{j_l})$, where $\bw=\bv_J/2$.

Having defined the odds ratio increment in \re{DeltasbvJ}, we are ready
to state the main result.
It tells that the odds ratio for exposure $\bv_J$ is a sum of the odds
ratio increments. This includes the baseline odds ratio when none of
the factors in $J$ are turned on, the additive marginal odds ratio
increments for those $|\bv_J|$ factors in $J$ that are exposed, and the
odds ratio interactions of order $2,3,\ldots,|\bv_J|$ among these
factors (see Appendix \ref{appa} for a proof):

\begin{theorem}\label{thm1}
Suppose $\bv_K$ is fixed. Then the odds ratios $\OR_{\sbv}$ admit
an additive expansion
\beq
\OR_{\sbv} = \OR_{(\sbv_J,\sbv_K)} = \sum_{\sbfm{0}\le\sbw\le\sbv_J}
\varDelta^{(\sbw,\sbv_K)} \OR
\lb{ORDecomp}
\eeq
for all $\bv_J$, with $\varDelta^{(\sbw,\sbv_K)} \OR$ as defined in \re
{DeltasbvJ}. Conversely, if \re{ORDecomp} holds and $\varDelta^{(\sbw,\sbv
_K)} \OR$ is a linear combination of $\{\OR_{(\sbu,\sbv_K)}; \, 0\le\bu
\le\bw\}$ for each $\bzero\le\bw\le\bv_J$, then necessarily $\varDelta
^{(\sbw,\sbv_K)} \OR$ must satisfy \re{DeltasbvJ}.
\end{theorem}

The expansions in \re{DeltasbvJ} and \re{ORDecomp} define a
combinatorial inclusion--exclusion principle. In order to see this, we
associate to each $j\in J$ a set $A_j$ and its complement $A_j^c$. Then
$\varDelta^{(\sbv_J,\sbv_K)} \OR$ represents an intersection $ (\xch{\bigcap
_{j; v_j=1}}{\cap
_{j; v_j=1}} A_j ) \cap (\xch{\bigcap_{j; v_j=0}}{\cap_{j; v_j=0}} A_j^c )$,
whereas $\OR_{(\sbv_J,\sbv_K)}$ corresponds to a union $\xch{\bigcup_{j;
v_j=1}}{\cup_{j;
v_j=1}}
A_j$, and in particular $\OR_{(\sbfm{0},\sbv_K)}=\emptyset$.

It is of interest to know how much of the odds ratio \re{OR}
%that
can be explained by marginal effects and lower order interaction among
the factors
in $J$. For this reason we introduce the following concept:

\begin{definition}
A prediction
\beq
\OR_{(\sbv_J,\sbv_K),i} = \sum_{\sbfm{0}\le\scbfm{w}\le\scbfm{v}_J\atop
|\scbfm{w}|\le i} \varDelta^{(\sbw,\sbv_K)} \OR
\lb{ORPred}
\eeq
of the odds ratio \re{OR} is obtained by including only terms up to
order $i$ in \re{ORDecomp}, for some $0\le i\le|\bv_J|$.
\end{definition}

It is possible to rewrite the predicted odds ratio \re{ORPred} as a
linear combination of lower order odds ratios $\{\OR_{(\sbw,\sbv_K)};
\, 0\le\bw\le\bv_J, |\bw|\le i\}$ for exposure vectors $\bw$ that have
at most $i$ factors in $J$ at risk (see Appendix \ref{appb} for a proof):

\begin{proposition}\label{prop1}
The prediction of the odds ratio $\OR_{(\sbv_J,\sbv_K)}$ in \re
{ORPred}, based on marginal effects among the factors in $J$ at risk
and their interaction terms up to order $i$, satisfies
\beq
\OR_{(\sbv_J,\sbv_K),i}
= \sum_{\sbfm{0}\le\scbfm{w}\le\scbfm{v}_J\atop|\scbfm{w}|\le i} \OR
_{(\sbw,\sbv_K)}
(-1)^{i-|\sbw|} {|\sbv_J|-1-|\sbw|\choose i-|\sbw|}
\lb{ORPred2}
\eeq
if $0\le i<|\bv_j|$, and
\beq
\OR_{(\sbv_J,\sbv_K),i} = \OR_{(\sbv_J,\sbv_K)}
\lb{ORPred3}
\eeq
if $i=|\bv_j|$.
\end{proposition}

Assume from now on that $\bv_J=\bone$, so that all factors in $J$ are
at risk. We will quantify how much of the
odds ratio $\OR_{(\sbfm{1},\sbv_K)}$
%that
is left unexplained by marginal effects and lower orders of interaction
among the factors in $J$. To this end, we introduce the following concept:

\begin{definition}
The unadjusted excess odds ratio
\beq
\UEOR_{(\sbfm{1},\sbv_K),i} = \OR_{(\sbfm{1},\sbv_K)} - \OR_{(\sbfm
{1},\sbv_K),i-1}
\lb{UEOR}
\eeq
is the difference between the odds ratio \re{OR} and a prediction \re
{ORPred} of it due to terms of order less than $i$, where $1\le i \le|J|$.
\end{definition}

The unadjusted excess odds ratio can be interpreted as an
unstandardized residual of a regression model, where only marginal
effects and interaction terms up to order $i-1$ are included as
independent variables in order to predict the odds ratio \re{OR}.
Equivalently, it quantifies the contribution of the odds ratio
expansion \re{ORDecomp} from terms of order at least $i$. The special
case $i=|J|$ is treated by Katsoulis and Bamia \cite{Kat17} when $K=\emptyset$.
The definition of the unadjusted odds ratio then simplifies to $\varDelta
^{(\sbfm{1},\sbv_K)} \OR$.

We will define three measures \re{xi} of marginal, joint or interaction
effects among the factors in $J$, and
all of them are functions of $\UEOR$.

\begin{definition}
The excess odds ratio
\begin{align}
\xi&= \EOR_{(\sbfm{1},\sbv_K),i}\notag\\[-1pt]
&= \frac{\sOR_{(\scbfm{1},\scbfm{v}_K)}}{\sOR_{(\scbfm{0},\scbfm
{v}_K)}} - \frac{\sOR_{(\scbfm{1},\scbfm{v}_K),i-1}}{\sOR_{(\sbfm
{0},\scbfm{v}_K)}}\notag\\[-1pt]
&= \frac{\sUEOR_{(\scbfm{1},\scbfm{v}_K),i}}{\sOR_{(\scbfm{0},\scbfm{v}_K)}}\xch{}{,}\lb{EOR}
\end{align}
expresses \re{UEOR} in units of the odds ratio when no factor in $J$ is
at risk, but those in $K$ are kept fixed at level $\bv_K$.
It has a range of $(-\infty,\infty)$, with $\EOR=0$ indicating absence
of effect.
\end{definition}

\begin{definition}
The quantity
\begin{align}
\xi&= \AP_{(\sbfm{1},\sbv_K),i}\notag\\[-1pt]
&= \frac{\sUEOR_{(\scbfm{1},\scbfm{v}_K),i}}{{\scr{max}}(\sOR_{(\scbfm
{1},\scbfm{v}_K)},\sOR_{(\scbfm{1},\scbfm{v}_K),i-1})}\notag\\[-1pt]
&= \frac{\sUEOR_{(\scbfm{1},\scbfm{v}_K),i}}{{\scr{max}}(\sOR_{(\scbfm
{1},\scbfm{v}_K)},\sOR_{(\scbfm{1},\scbfm{v}_K)}-\sUEOR_{(\scbfm
{1},\scbfm{v}_K),i})}\lb{AP}
\end{align}
is the attributable proportion of the odds ratio due to terms \re
{DeltasbvJ} of order at least $i$.
It uses the same denominator as in H\"{o}ssjer et al. \cite{Hos17} in order
to assure $-1\le\AP\le1$, with $\AP=0$ indicating absence of effect.
\end{definition}

\begin{definition}
The synergy index
\begin{align}
\xi&= \SI_{(\sbfm{1},\sbv_I),i}\notag\\[-1pt]
&= \frac{\sOR_{(\scbfm{1},\scbfm{v}_K)}-\sOR_{(\scbfm{0},\scbfm
{v}_K)}}{\sOR_{(\scbfm{1},\scbfm{v}_K),i-1}-\sOR_{(\scbfm{0},\scbfm
{v}_K)}}\notag\\
&= \frac{\sOR_{(\scbfm{1},\scbfm{v}_K)}-\sOR_{(\scbfm{0},\scbfm
{v}_K)}}{\sOR_{(\scbfm{1},\scbfm{v}_K)}-\sUEOR_{(\scbfm{1},\scbfm
{v}_K),i}-\sOR_{(\scbfm{0},\scbfm{v}_K)}}\xch{}{,}
\lb{SI}
\end{align}
is only defined in order to quantify interaction ($2\le i \le|J|$),
since otherwise the denominator of \re{SI} vanishes. It is also
required that the joint and lower order
effects of the factors in $J$ %is
are positive ($\OR_{(\sbfm{1},\sbv_K)}>\OR_{(\sbfm{0},\sbv_K)}$ and $\OR
_{(\sbfm{1},\sbv_K),i-1}>\OR_{(\sbfm{0},\sbv_K)}$) in order for SI to
be meaningful. Then its range is $(0,\infty)$, with $\SI=1$ indicating
absence of effect.
\end{definition}

All three quantities in equations \re{EOR}--\re{SI} are stratified for the
risk factors in $K$, like confounders that are controlled at their
observed levels. In contrast,
there is only partial (additive) control for the remaining $q$
covariates of $\bz$. EOR and AP can be viewed as different types of
standardized residuals of a regression model where marginal effects and
lower order interactions among the factors in $J$ are used to predict
the odds ratio \re{OR}.
The inverse synergy index $\SI^{-1}$ is the analogue of the coefficient
of determination. It quantifies how
large fraction of the odds ratio increment above the baseline
level $\OR_{(\sbfm{0},\sbfm{v}_K)}$ that is explained by the regression model.

Some special cases of formulas \re{EOR}--\re{SI} are of particular
interest. When $i=|J|=1$, there is one single risk factor ($J=\{1\}$).
Equation \re{EOR} then quantifies
the excess odds ratio due the marginal effect (EORM) of factor 1, when
those in $K$ are controlled for at level $\bv_K$.
Similarly, equation \re{AP} defines the attributable proportion of risk
due to the marginal effect of factor 1, whereas the synergy index is
not well defined.

When $i=1$ and $|J|\ge2$, we refer to the quantity in \re{EOR} as
EORJ, the excess odds ratio due to joint (marginal and interaction)
effects of all factors in $J$ when those in $K$ are controlled at level
$\bv_K$. In the same way, $\AP$ is the attributable proportion due to
joint effects among the factors in $J$, whereas the synergy index is undefined.

When $i=2$ we refer to the quantity in \re{EOR} as the total excess
odds ratio due to all levels of interaction (TotEORI) among the factors
in $J$, when those in $K$ are controlled at level $\bv_K$. $\AP$ and
$\SI$ are similarly referred to as the attributable proportion and
synergy index due to all levels of interaction among the factors in $J$.

Finally, when $i=|J|$ we refer to the quantity in \re{EOR} as the
excess odds ratio due to the highest order $|J|$ of interaction (EORI)
among the factors in $J$, when those in $K$ are controlled at level $\bv
_K$. Analogously, equations \re{AP}--\re{SI} quantify the attributable
proportion and the synergy index due to the highest order $|J|$ of
interaction among the factors in $J$.

Since EOR, AP and SI are all functions of UnadjEOR, it suffices to
specify the latter. In the subsections to follow we will do so for
models with 1, 2, or 3 risk factors in $J$.

\subsection{One risk factor}

When $|J|=1$ there is only one possible unadjusted excess odds ratio
\[
\UEOR_{(1,\sbv_K),1} = \OR_{(1,\sbv_K)} - \OR_{(0,\sbv_K)},
\]
caused by the marginal effect of factor 1.

\subsection{Two risk factors}

When $|J|=2$ there are only two possible unadjusted excess odds ratios
\[
\UEOR_{(1,1,\sbv_K),i} \,{=}
\begin{cases}
\OR_{(1,1,\sbv_K)}\,{-}\,\OR_{(0,0,\sbv_K)}, & i\,{=}\,1,\\
\OR_{(1,1,\sbv_K)}\,{-}\,\OR_{(1,0,\sbv_K)}\,{-}\,\OR_{(0,1,\sbv_K)}\,{+}\,\OR_{(0,0,\sbv_K)}, & i\,{=}\,2,
\end{cases}
\]
obtained by inserting \re{ORPred} into \re{UEOR}. The first unadjusted
excess odds ratio is due to the joint (marginal and interaction) effect
of factors 1 and 2, whereas the second only includes interaction
between these two factors.

\subsection{Three risk factors}

When $|J|=3$ there are three possible unadjusted excess odds ratios
\[
\UEOR_{(1,1,1,\sbv_K),i} =
\begin{cases}
\OR_{(1,1,1,\sbv_K)}-\OR_{(0,0,0,\sbv_K)}, & i=1,\\
\OR_{(1,1,1,\sbv_K)}-\OR_{(1,0,0,\sbv_K)} & \\
\quad -\,\OR_{(0,1,0,\sbv_K)} -\OR_{(0,0,1,\sbv_K)} & \\
\quad +2\,\OR_{(0,0,0,\sbv_K)}, & i=2,\\
\OR_{(1,1,1,\sbv_K)}-\OR_{(1,1,0,\sbv_K)} & \\
\quad -\,\OR_{(1,0,1,\sbv_K)} -\OR_{(0,1,1,\sbv_K)} & \\
\quad +\,\OR_{(1,0,0,\sbv_K)}+\OR_{(0,1,0,\sbv_K)} & \\
\quad +\,\OR_{(0,0,1,\sbv_K)}-\OR_{(0,0,0,\sbv_K)}, & i=3,\\
\end{cases}
\]
all of them derived by inserting \re{ORPred} into \re{UEOR}. The first
unadjusted excess odds ratio is caused by the joint (marginal and
interaction) effect of factors 1, 2 and 3, the second includes second
and third order interaction between these three factors but no marginal
effects \cite{Pet13,Hos17}, whereas the
third only includes the highest third order interaction between factors
1, 2 and 3 \cite{Kat17}.

\section{Inference of joint effects and interaction}\lb{Sec:Inf}

Assume that a case-control dataset $(\bx_1,Y_1),\ldots,(\bx_n,Y_n)$ of
size $n=n_0+n_1$ is available, with $n_0$ controls ($Y_a=0$) and $n_1$
cases ($Y_a=1$). Since this is a retrospective sample, we cannot
estimate all parameters of the logistic regression model \re{etax}. On
the other hand, it is possible to estimate the structural parameters
$\bpsi$ consistently when controls are frequency matched with cases.
This is the unconditional logistic regression approach, whereby the
prospective log likelihood
\beq
(\hbpsi,\hat{\bka}) = \arg\max_{\sbpsi,\sbka} L(\bpsi,\bka) = \arg\max_{\sbpsi,\sbka} \sum_{a=1}^n \log P(Y_a|\bx_a)
\lb{MLE}
\eeq
is maximized over the model parameters in \re{etax}. This approach can
also be used if cases and controls are matched in strata, as long as
all variables used for matching are included in the model as
covariates, and the number of strata does not grow with sample size. On
the other hand, conditional logistic regression is more appropriate if
the number of strata is large \cite{Bre80}.

Let
\beq
\hxi= \xi(\hbpsi)
\lb{hxi}
\eeq
be the plug-in estimator of \re{xi} obtained from the maximum
likelihood estimator \re{MLE}.
In order to produce a confidence interval for $\xi$ with asymptotic
coverage probability $1-\alpha$ we use
the delta method in conjunction with some appropriately chosen monotone
increasing and differentiable transformation $h$. This leads to
\beq
\CI=  \bigl[ h^{-1} \bigl(h(\hxi)-z_{1-\alpha/2}\SE \bigr),h^{-1}
\bigl(h(\hxi)+z_{1-\alpha/2}\SE \bigr) \bigr],
\lb{CI}
\eeq
where $z_\beta$ is the $\beta$-quantile of a standard normal
distribution and $\SE$ is the standard error of $h(\hxi)$.
This method relies on asymptotic normality
\beq
\hbpsi\sim\mbox{AsN}(\bpsi,\bSi)
\lb{AsN}
\eeq
of the structural part of \re{MLE} for large samples,
with $\bSi=\bI(\bpsi,\bka)^{-1}$ an asymptotic approximation of the
covariance matrix $\Cov(\hbpsi)$ of $\hbpsi$, which equals the inverse
of the Fisher information matrix $\bI$ \cite{Cas02}. In
order to find $\SE$ we first approximate the asymptotic variance of
$\hxi$ by $\si^2 = \bD\bSi\bD^T$, using \re{AsN} and
%a
the first order Taylor expansion of $\xi(\bpsi)$, with $\bD= \bD(\bpsi
)=d\xi(\bpsi)/d\bpsi$ and $T$ referring to vector transposition (see
Appendix \ref{appc} for an explicit expression of $\bD$). Then one chooses the
function $h$ so that the distribution of $h(\hxi)$ is closer to normal
than that of $\hxi$. Typically $h$ is a variance stabilizing
transformation that
maps the range of $\xi$ to the real line $(-\infty,\infty)$.
%A
The Taylor expansion of $h$ gives the asymptotic variance of $h(\hxi)$.
By taking the square root of an estimate of this variance,
we find the standard error
\[
\SE= \sqrt{\widehat{\Var}\bigl(h(\hxi)\bigr)} = h^\prime(\hxi)\hsi
\]
of $h(\hxi)$, with $\hsi^2 = \hat{\bD}\hat{\bSi}\hat{\bD}^T$ an
estimate of $\si^2$, $\hat{\bD}=\bD(\hat{\bpsi})$ an estimate of $\bD$,
and $\hat{\bSi}=\bI(\hbpsi,\hbka)^{-1}$ the inverse of the observed
Fisher information matrix. In this paper we will use the transformations
\beq
h(\xi) =
\begin{cases}
\xi, & \mbox{if }\xi=\EOR,\\
\log \bigl[(1+\xi)/(1-\xi) \bigr], & \mbox{if }\xi=\AP,\\
\log(\xi), & \mbox{if }\xi=\SI,
\end{cases}
\lb{h}
\eeq
see for instance Rothman \cite{Rot76} and H\"{o}ssjer et al. \cite{Hos17}. The
three functions in \re{h}
map the corresponding indices $\xi$ from their original ranges
($(-\infty,\infty)$ for EOR, $(-1,1)$ for AP, $(0,\infty)$ for SI) to
$(-\infty,\infty)$.

\section{Analysis of a real dataset}\lb{Sec:Data}

Multiple sclerosis (MS) is a complex and inflammatory disease causing
damage to the central nervous system. Its prevalence is over 0.1\% in
many countries, affecting large regions of the world
\cite{MSI13}. There is solid evidence for a genetic component of
the disorder, with a major contribution from variants at the human
leukocyte antigen (HLA) complex. It is also well known that presence of
allele 15 of the HLA-DRB1 gene is a risk factor, whereas allele 02 of
the HLA-A gene has a protective effect \cite{Cre14}. Several studies
reveal that environmental factors, in particular smoking, impact the
risk of the disease as well \cite{Ols17}. A Swedish study in
Hedstr\"{o}m et al. \cite{Hed11} demonstrated positive pairwise additive
interaction between the two genetic factors, and also between smoking
and each genetic factor. These results have more recently been
replicated and refined in Hedstr\"{o}m et al. \cite{Hed17}, by merging
case-control studies from several countries.
%Because of
Due to the size of this meta analysis, it was also possible to
investigate whether third order interaction was present between the two
genetic factors and smoking. In order to illustrate the methodology of
this paper, we present some of the findings from the four Nordic
studies (see Table~\ref{MSStudies}) of Hedstr\"{o}m et al. \cite{Hed17}.

\begin{table}
\caption{Number of cases and controls}\lb{MSStudies}%
\begin{tabular}{lD{.}{.}{4.0}D{.}{.}{4.2}}\hline
Study & \multicolumn{1}{l}{Cases} & \multicolumn{1}{l}{Controls}\\ \hline
Swedish EIMS study & 1308 & 1858\\
Swedish GEMS study & 3272 & 2382\\
Danish study & 1474 & 3469\\\smallskip
Norwegian study & 211 & 692\\
Combined Nordic study & 6265 & 8401\\\hline \\
\multicolumn{3}{L{140pt}}{The four Nordic studies are from Hedstr\"{o}m et
al. \cite{Hed17}}
\end{tabular}
\end{table}

Apart from the two genetic factors and smoking, three other covariates
(gender, age, study) were also part of the model. This gives a total of
8 covariates, encoded as
\begin{align*}
x_1 &= \mbox{HLA-DRB1 (1 for genotypes with a least one copy of},\\
&\quad\ \ \mbox{allele 15, 0 otherwise)}, \\
x_2 &= \mbox{HLA-A (1 for genotypes with no allele 02, 0 otherwise)},
\\
x_3 &= \mbox{smoking (1$=$smoker, 0$=$non-smoker)},\\
x_4 &= \mbox{gender (1$=$female, 0$=$male)},\\
x_5 &= \mbox{age when MS was detected \big($\in\{0,1,\ldots,73\}$\big)},\\
x_6-x_8 &= \mbox{study \big((0,0,0)$=$EIMS, (1,0,0)$=$GEMS, (0,1,0)$=$Danish,}\\
&\quad\ \, \mbox{(0,0,1)$=$Norwegian\big)}.
\end{align*}
The last three study covariates were only included for the meta
analysis, so that
\[
p+q =
\begin{cases}
5, & \mbox{for each separate study},\\
8, & \mbox{for the combined Nordic study}.
\end{cases}
\]
Let
\beq
\xi= \frac{\OR_{(\sbfm{1},\sbv_K)}}{\OR_{(\sbfm{0},\sbv_K)}}
\lb{xiOR}
\eeq
be the odds ratio for the joint effect of all the risk factors $J$,
when the confounding risk factors in $K$ are fixed at level $\bv_K$,
and the remaining covariates are $\{1,\ldots,\break p+q\}\setminus(J\cup K)$.
Table~\ref{Tab:OR} lists maximum likelihood estimates \re{hxi} of these
odds ratios for various choices of risk factors and confounders.
The associated confidence intervals \re{CI} are obtained using the
logarithmic transformation in \re{h}. Although\vadjust{\eject} MS is a common disorder,
its prevalence is
small enough to assume that the point estimates and confidence
intervals of Table~\ref{Tab:OR} are representative for the relative
risks $\RR_{(\sbfm{1},\sbv_K,\sbfm{z})}/\RR_{(\sbfm{0},\sbv_K,\sbfm
{z})}$ as well.

We find, for instance, that the point estimate of the marginal odds
ratio (or relative risk) of having MS in the combined dataset is 3.6
for individuals with the DRB15 risk allele, compared to those that lack
this allele.
The corresponding marginal odds ratios for absence of the protecting A2
allele and for smoking are 1.75 and 2.0 respectively. Since the joint
odds ratios for all pairs of risk factors are much larger than the
corresponding marginal odds ratios, there are strong indications of
two-way interactions between all pairs of risk factors. There is
possibly some three-way interaction between DR15, A2- and smoking as
well, since the joint OR for all three factors is higher than the
pairwise odds ratios. On the other hand, the OR for the two genetic
factors is only higher among smokers than among non-smokers for one
study (EIMS).

\begin{table}[t!]
\caption{Point estimates and 95\% confidence intervals of the odds
ratio \re{xiOR}}\lb{Tab:OR}
\begin{tabular}{lD{.}{.}{2.14}D{.}{.}{2.14}D{.}{.}{2.14}@{}}\hline
Study & \multicolumn{3}{l}{OR for one factor} \\
\cline{2-4}
& \multicolumn{1}{l}{DR15} & \multicolumn{1}{l}{A2-} & \multicolumn{1}{l}{sm} \\
\hline
EIMS & 3.55\ (3.05,4.13) & 1.74\ (1.50,2.02) & 1.52\ (1.30,1.78)\\
GEMS & 3.70\ (3.30,4.15) & 1.79\ (1.60,2.00) & 1.62\ (1.44,1.82)\\
Danish & 3.42\ (2.99,3.92) & 1.73\ (1.51,1.98) & 3.09\ (2.70,3.55)\\\smallskip
Norwegian & 5.02\ (3.50,7.21) & 1.77\ (1.24,2.53) & 2.13\ (1.50,3.04)\\
Combined & 3.60\ (3.34,3.87) & 1.75\ (1.63,1.88) & 2.00\ (1.86,2.15)\\
\hline
Study & \multicolumn{3}{l}{OR for two factors} \\
\cline{2-4}
& \multicolumn{1}{l}{DR15, sm} & \multicolumn{1}{l}{A2-, sm} & \multicolumn{1}{l}{DR15, A2-} \\
\hline
EIMS & 5.41\ (4.29,6.85) & 2.71\ (2.17,3.40) & 6.18\ (4.94,7.75)\\
GEMS & 5.83\ (4.93,6.91) & 2.89\ (2.45,3.42) & 6.44\ (5.46,7.60)\\
Danish & 10.49\ (8.57,12.84) & 5.35\ (4.38,6.52) & 5.95\ (4.89,7.23)\\ \smallskip
Norwegian & 10.86\ (6.53,18.07) & 3.78\ (2.27,6.27) & 8.84\ (5.21,14.99)\\
Combined & 7.11\ (6.37,7.93) & 3.51\ (3.16,3.91) & 6.28\ (5.64,6.98)\\
\hline
Study & \multicolumn{3}{l}{OR for three factors/confounders} \\
\cline{2-4}
& \multicolumn{1}{l}{DR15, A2-|nsm} & \multicolumn{1}{l}{DR15, A2-|sm} & \multicolumn{1}{l}{DR15, A2-, sm}\\
\hline
EIMS & 5.62\ (4.27,7.39) & 7.72\ (5.17,11.52) & 11.23\ (7.81,16.14)\\
GEMS & 7.07\ (5.70,8.77) & 5.61\ (4.34,7.27) & 9.96\ (7.75,12.79)\\
Danish & 6.57\ (5.01,8.61) & 5.27\ (3.95,7.01) & 17.95\ (13.34,24.17)\\ \smallskip
Norwegian & 9.02\ (4.29,18.98) & 8.66\ (4.14,18.11) & 18.27\ (8.62,38.72)\\
Combined & 6.37\ (5.55,7.31) & 6.16\ (5.20,7.29) & 12.63\ (10.73,14.85)\\
\hline
\multicolumn{4}{c}{}\\
\multicolumn{4}{L{271pt}@{}}{The sets of risk factors is $J$ and confounders
fully controlled for is $K$. Each column is denoted as $J$ when $K=\emptyset$, and
otherwise as $J|K$. The confidence intervals are given in brackets. We use the notation
DR15 for presence of allele 15 at HLA-DRB1, A2-for absence of allele 2 at
HLA-A, sm for smoker and nsm for non-smoker}
\end{tabular}    %\vspace*{-4pt}
\end{table}

The estimates of Table~\ref{Tab:OR} motivate further analysis of the MS
datasets in Hedstr\"{o}m et al \cite{Hed17} based on the attributable
proportion, excess odds ratio and synergy index for the three risk
factors DR15, A2- and smoking.
Table~\ref{APEORSI:ThreeFac} gives confidence intervals for all three
quantities when $J=\{1,2,3\}$ and $K=\emptyset$, for various choices of
$i$. It confirms a strong joint effect of all three factors, since AP
and EOR are both significantly different from 0, and SI is
significantly different from 1.
For instance, the estimate $\widehat{\AP}=0.92$ for the combined Nordic
data set indicates that about 92\% of the odds ratio (or disease risk)
for smokers with both genetic risk factors, is not present among those
that lack all three factors. We also find that the total amount of
second and third order interaction between DR15, A2- and smoking is
strongly significant. In particular, $\widehat{\AP}=0.52$ for the
Nordic metapopulation indicates that about half of the disease risk is
due to interaction. One the other hand, $\widehat{\SI}= 2.38$ tells
that the disease risk increment of a smoker with both genetic risk
factors, compared to an individual with none of these risk factors, is
more than twice the disease risk increment due to marginal effects only.
Finally, we find from the rightmost AP, EOR and SI columns of Table~\ref{APEORSI:ThreeFac}
that third order interaction between DR15, A2- and
smoking is only significant for one dataset (EIMS).

We have also estimated the attributable proportion separately for males
and females (data not shown). The results are in quite good agreement
with the upper part of
Table~\ref{APEORSI:ThreeFac}, although the values for males are
slightly larger than those for females, and since the gender specific
datasets are smaller, the confidence intervals are wider. Such an
analysis illustrates how joint effects of and interaction among three
factors ($J=\{1,2,3\}$) is affected when one controls for a fourth
factor ($K=\{4\}$).

\begin{table}[t!]
\caption{Point estimates and 95\% confidence intervals for AP, EOR and
SI}\lb{APEORSI:ThreeFac}
\begin{tabular}{lD{.}{.}{2.14}D{.}{.}{2.14}D{.}{.}{2.15}@{}}\hline
Study & \multicolumn{3}{l}{AP for DR15, A2-, smoking} \\
\cline{2-4}
& \multicolumn{1}{l}{\xch{Joint}{joint} effects} & \multicolumn{1}{l}{2nd \& 3rd order interaction} & \multicolumn{1}{l}{3rd order interaction}\\
\hline
EIMS & 0.91\ (0.87,0.94) & 0.59\ (0.43,0.72) & 0.39\ (0.12,0.61) \\
GEMS & 0.90\ (0.87,0.92) & 0.35\ (0.19,0.50) & -0.02\ (-0.29,0.25) \\
Danish & 0.94\ (0.93,0.96) & 0.60\ (0.49,0.70) & 0.09\ (-0.18,0.34) \\ \smallskip
Norwegian & 0.95\ (0.89,0.97) & 0.66\ (0.36,0.84) & 0.22\ (-0.33,0.66)\\
Combined & 0.92\ (0.91,0.93) & 0.53\ (0.46,0.60) & 0.15\ (0.00,0.29)\\
\hline
Study & \multicolumn{3}{l}{EOR for DR15, A2-, smoking} \\
\cline{2-4}
& \multicolumn{1}{l}{\xch{Joint}{joint} effects} & \multicolumn{1}{l}{2nd \& 3rd order interaction} & \multicolumn{1}{l}{3rd order interaction}\\
\hline
EIMS & 10.23\ (6.15,14.30) & 6.68\ (3.05,10.31) & 4.35\ (0.33,8.37)\\
GEMS & 8.96\ (6.46,11.45) & 3.50\ (1.32,5.68) & -0.24\ (-3.05,2.58)\\
Danish & 16.95\ (11.62,22.29) & 10.85\ (6.55,15.14) & 1.55\ (-3.41,6.51)\\\smallskip
Norwegian & 17.27\ (3.55,30.99) & 12.05\ (0.82,23.28) & 4.07\
(-7.82,15.95)\\
Combined & 11.63\ (9.57,13.68) & 6.74\ (5.00,8.49) & 1.88\ (-0.16,3.93)\\
\hline
Study & \multicolumn{3}{l}{SI for DR15, A2-, smoking} \\
\cline{2-4}
& \multicolumn{1}{l}{\xch{Joint}{joint} effects} & \multicolumn{1}{l}{2nd \& 3rd order interaction} & \multicolumn{1}{l}{3rd order interaction}\\
\hline
EIMS & \multicolumn{1}{l}{undefined} & 2.88\ (1.87,4.43) & 1.74\ (1.10,2.75)\\
GEMS & \multicolumn{1}{l}{undefined} & 1.64\ (1.25,2.16) & 0.97\ (0.71,1.33)\\
Danish & \multicolumn{1}{l}{undefined} & 2.78\ (2.07,3.72) & 1.10\ (0.81,1.49)\\  \smallskip
Norwegian & \multicolumn{1}{l}{undefined} & 3.31\ (1.50,7.32) & 1.31\ (0.62,2.76)\\
Combined & \multicolumn{1}{l}{undefined} & 2.38\ (2.00,2.84) & 1.19\ (0.99,1.44)\\
\hline
\multicolumn{4}{c}{}\\
\multicolumn{4}{l}{The measures either quantify joint marginal and
interaction effects ($i=1$), the total inter-}\\
\multicolumn{4}{l}{action of order 2 and 3 ($i=2$), or the highest
order 3 of interaction ($i=3$) between the}\\
\multicolumn{4}{l}{three risk factors DR15, A2- and smoking ($J=\{
1,2,3\}$). No other covariates are fully }\\
\multicolumn{4}{l}{controlled for (so that $p=3$ and$K=\emptyset$). The
confidence intervals are given in brackets.}\\
\multicolumn{4}{l}{Notice that SI is only defined for measures of interaction}
\end{tabular}
\end{table}
%
%\par\medskip
%\begin{table}
%\begin{center}
%{
%\begin{tabular}{|c|c|c|c|}
%\hline
%Study & \multicolumn{3}{c|}{AP for DR15,A2-,smoking among males} \\
%\cline{2-4}
%& joint effects & 2nd \& 3rd order interaction & 3rd order interaction\
%\
%\hline
%EIMS & 0.92 (0.84,0.96) & 0.58 (0.20,0.81) & 0.36 (-0.22,0.75)\\
%GEMS & 0.91 (0.85,0.95) & 0.36 (0.01,0.63) & 0.21 (-0.29,0.62)\\
%Danish & 0.96 (0.94,0.98) & 0.69 (0.54,0.79) & 0.23 (-0.13,0.53)\\
%Norwegian & 0.98 (0.90,1.00) & 0.84 (0.46,0.96) & 0.55 (-0.22,0.90)\\
%\hline
%Combined & 0.94 (0.91,0.95) & 0.58 (0.46,0.69) & 0.25 (0.01,0.46)\\
%\hline
%\hline
%Study & \multicolumn{3}{c|}{AP for DR15,A2-,smoking among females} \\
%\cline{2-4}
%& joint effects & 2nd \& 3rd order interaction & 3rd order interaction\
%\
%\hline
%EIMS & 0.91 (0.86,0.94) & 0.60 (0.40,0.74) & 0.38 (0.06,0.63)\\
%GEMS & 0.90 (0.86,0.92) & 0.35 (0.16,0.52) & -0.10 (-0.37,0.18)\\
%Danish & 0.93 (0.90,0.95) & 0.51 (0.33,0.65) & -0.05 (-0.38,0.29)\\
%Norwegian & 0.92 (0.82,0.97) & 0.54 (0.09,0.81) & 0.04 (-0.65,0.69)\\
%\hline
%Combined & 0.91 (0.89,0.93) & 0.50 (0.40,0.59) & 0.08 (-0.11,0.26)\\
%\hline
%\multicolumn{4}{c}{}\\
%\end{tabular}
%}
%\\
%\caption{Point estimates for AP due to joint marginal and interaction
%effects ($i=1$), the total interaction of order 2 and 3 ($i=2$), and
%the highest order 3 of interaction ($i=3$) between the three risk
%factors DR15, A2- and smoking ($J=\{1,2,3\}$), when gender is fully
%controlled for (so that $p=4$ and $K=\{4\}$). The associated 95\%
%confidence intervals are given in brackets.}\lb{AP:ThreeFacGend}
%\end{center}
%\end{table}

\section{Discussion}\lb{Sec:Disc}

In this paper we studied how a collection $J$ of binary risk factors
jointly influence a binary outcome. Our objective was to develop a
general framework for quantifying marginal effects and various orders
of interaction between these factors on an additive scale, when
stratifying or partially controlling for other confounding variables.
This led to the key result; an expansion of the odds ratios as a sum of
marginal effects and interaction terms of different orders, with a
finite difference and a combinatorial inclusion--exclusion
interpretation. We also showed how to use these odds ratio expansions
for estimating and producing confidence intervals for the excess odds
ratio, attributable proportion and synergy index from a case-control
dataset. The methodology was illustrated using a Nordic meta study for
multiple sclerosis. The inferential procedure relies on maximum
likelihood estimates from a logistic regression model and the delta
method. It has been implemented in a Matlab program that is available
from the first author upon request.

Our approach makes it possible to stratify for some variables (in $K$)
at various levels, and yet use the whole data set to estimate effect
parameters of the other $q$ confounders, for which there is only
partial control. But this requires a large data set in order to
estimate all $2^{p}+q=2^{|J|+|K|}+q$ parameters of the logistic
regression model. An alternative and simpler strategy is to use only
those observations $(\bx_a,Y_a)$ for which
the variables in $K$ are at a pre-specified level. This gives a smaller
model with only $2^{|J|}+q$ parameters to estimate.

The delta based confidence intervals are fast to compute. This is
suitable in applications where a number of different putative risk
factors are sought for. We have implemented confidence intervals based
on resampling as well, using the bias-corrected accelerated percentile
method \cite{Efr87,Efr94,Hos17}. There is generally good agreement between the resampling and
delta-based intervals for odds ratios and attributable proportions, as
long as the interaction effects are not too strong, the order of
interaction is not too high and the data set is not too small. On the
other hand, the delta based confidence intervals tend to be less
accurate for excess odds ratios. For this reason we recommend to report
excess odds ratios with resampling based confidence intervals. As a
topic of future research,
it would be of interest to compare the asymptotic accuracy of the delta
and resampling based confidence intervals more systematically.

Another object of further study is to develop odds ratio expansions of
main effects and interactions when some of the risk factors are
continuous \cite{Eid01} and to find analogous expansions for non-binary
outcomes.

\section*{Acknowledgement}

The authors wish to thank the editors and three anonymous reviewers for
helpful comments that considerably improved the structure and content
of the paper. Ola H\"{o}ssjer was financially supported by the Swedish
Research Council, contract \xch{Nr.}{nr.}~621-2013-4633.

\begin{appendix}

\section{Proof of Theorem \ref{thm1}}\label{appa}

Assume without loss of generality that $J=\{1,\ldots,p\}$ and
$K=\emptyset$. It is possible then to rewrite \re{ORDecomp} as
\beq
\varDelta^{\sbv} \OR= \OR_{\sbv} - \sum_{\sbfm{0}\le\sbw< \sbv} \varDelta
^{\sbw} \OR.
\lb{ORDecomp2}
\eeq
Since $\Delta^{\sbfm{0}}=1$, and the set $\{0,1\}^p$ of binary vectors
of length $p$ is partially ordered, we can use \re{ORDecomp} to compute
$\varDelta^{\sbv} \OR$ recursively for all $\bv\in\{0,1\}^p$. Such a
procedure gives
\beq
\varDelta^{\sbv} \OR= \sum_{\sbfm{0}\le\sbw\le\sbv} c_{\sbv,\sbw}\OR_{\sbw},
\lb{DeltasbvOR}
\eeq
for some constants $c_{\sbv,\sbw}$. In order to verify \re{DeltasbvJ}
and \re{ORDecomp}, it suffices to prove that
\beq
c_{\sbv,\sbw} = (-1)^{|\sbv|-|\sbw|}, \quad\mbox{for all }\bzero\le\bw
\le\bv.
\lb{c}
\eeq
We do this by induction with respect to $\bv\in\{0,1\}^p$. Starting
with $\bv=\bzero$, we notice first that $c_{\sbfm{0},\sbfm{0}}=1$, since
$\varDelta^{\sbfm{0}} \OR= \OR_{\sbfm{0}} = 1$. As a next step, consider
a fixed $\bv>\sbfm{0}$. In order to establish \re{c} for all $\bw$ such
that $\bzero\le\bw\le\bv$, we use the induction hypothesis and assume
that \re{c} holds for all ($\bw,\bu$) such that $\bzero\le\bw\le\bu<
\bv$. Then we insert \re{DeltasbvOR}--\re{c} into each term on the right
hand side of \re{ORDecomp2}, and change order of summation. This gives
\begin{align}
\varDelta^{\sbv} \OR&= \OR_{\sbv} - \sum_{\sbu;\sbfm{0}\le\sbu< \sbv}
\varDelta^{\sbu} \OR\notag\\
&= \OR_{\sbv} - \sum_{\sbu;\sbfm{0}\le\sbu< \sbv} \sum_{\sbw;\sbfm
{0}\le\sbw\le\sbu} (-1)^{|\sbu|-|\sbw|}\OR_{\sbw}\notag\\
&= \OR_{\sbv} - \sum_{\sbw;\sbfm{0}\le\sbw< \sbv} \OR_{\sbw} \sum_{\sbu;\sbw\le\sbu<\sbv} (-1)^{|\sbu|-|\sbw|}.
\lb{Exp}
\end{align}
It is clear from this expansion that $c_{\sbv,\sbv}=1$, so that \re{c}
holds for $\bw=\bv$. When $\bw<\bv$, we identify the coefficient of
$\OR_{\sbw}$ as
\begin{align*}
c_{\sbv,\sbw} &= -\sum_{\sbu;\sbw\le\sbu<\sbv} (-1)^{|\sbu|-|\sbw|}\\
&= -\sum_{k=0}^{|\sbv|-|\sbw|-1} (-1)^{k} \big|\big\{\bu;|\bu|=|\bw|+k,\bw\le
\bu\le\bv\big\}\big|\\
&= -\sum_{k=0}^{|\sbv|-|\sbw|-1} (-1)^{k} {|\sbv|-|\sbw|\choose k}\\
&= -  \bigl[(1-1)^{|\sbv|-|\sbw|}-(-1)^{|\sbv|-|\sbw|} \bigr]\\
&= (-1)^{|\sbv|-|\sbw|},
\end{align*}
and this finishes the proof of \re{c}.

\section{Proof of Proposition \ref{prop1}}\label{appb}

The second part of Proposition \ref{prop1}, equation \re{ORPred3}, concerns the
case when all orders of interaction are included for predicting the
odds ratio, i.e.\ $i=|\bv_J|$.
This equation follows directly from the fact that the definition of $\OR
_{(\sbv_J,\sbv_K),i}$ in \re{ORPred} is identical to \re{ORDecomp} when
$i=|\bv_J|$.

In order to prove the first part \re{ORPred2} of Proposition \ref{prop1}, when
$0\le i < |\bv_J|$, we insert the definition of $\varDelta^{(\sbw,\sbv
_K)}\OR$ in \re{DeltasbvJ} into equation \re{ORPred}. Then we change
the order of summation. This leads to
\beq
\OR_{(\sbv_J,\sbv_K),i}
= \sum_{\sbfm{0}\le\scbfm{w}\le\scbfm{v}_J\atop|\scbfm{w}|\le i} \OR
_{(\sbw,\sbv_K)} \sum_{l=0}^{i-|\sbw|} (-1)^l {|\sbv_J|-|\sbw|\choose l}.
\lb{ORPredX}
\eeq
Put $n=|\bv_J|-|\bw|$ and $m=i-|\bw|$, so that $0\le m < n$. The
combinatorical identity
\[
\sum_{l=0}^{m} (-1)^l
{n\choose l} = (-1)^{m}{n-1\choose m}
\]
can be proved by induction with respect to $m=0,1,\ldots,n-1$, using
Pascal's triangle in each step.
It implies that \re{ORPredX} is equivalent to \re{ORPred2}, and this
completes the proof.

\section{Expressions for $\bD(\bpsi)$}\label{appc}

It will be convenient to introduce the notation
\begin{align}
a &= \OR_{(\sbfm{1},\sbv_K)},\notag\\
b &= \OR_{(\sbfm{1},\sbv_K),i-1},\notag\\
c &= \OR_{(\sbfm{0},\sbv_K)},
\lb{abc}
\end{align}
respectively for the odds ratio when all the risk factors in $J$ are
turned on, the prediction of this odds ratio including main effects and
interaction terms up to order $i-1$,
and the odds ratio when all the risk factors in $J$ are turned off.
The excess odds ratio \re{EOR}, the attributable proportion \re{AP},
the synergy index \re{SI} and the adjusted odds ratio \re{xiOR} are all
different functions
\beq
\xi=
\begin{cases}
(a-b)/c, & \xi=\mbox{EOR},\\
(a-b)/\max(a,b), & \xi=\mbox{AP},\\
(a-c)/(b-c), & \xi=\mbox{SI},\\
a/c, & \xi=\mbox{adjusted OR},
\end{cases}
\lb{xiDef}
\eeq
of $a$, $b$ and $c$. Hence, in order to find the derivative $\bD$ of
$\xi$ with respect to $\bpsi$, for any of these quantities, it suffices
to find the derivatives of $a$, $b$ and $c$ with respect to $\bpsi$. To
this end, it is convenient to associate with any binary vector $\bu$ of
length $p$, another binary vector
\beq
\bone_{\le\sbu} = (1_{\sbw\le\sbu};\, \bfm{0}<\bw\le\bfm{1})
\lb{boneu}
\eeq
of length $2^p-1$. As in the definition of $\bpsi$ in \re{bpsi}, the
indices of $\bone_{\le\sbu}$ run over all binary vectors $\bfm{0}<\bw\le
\bfm{1}$ of length $p$ except the zero vector. For a given $\bw$, the
corresponding coordinate of $\bone_{\le\sbu}$ is 1 if $\bw\le\bu$ and
zero otherwise.

With these preliminaries, we can state the following result, which
follows from \re{abc}--\re{xiDef} and the definition of the odds ratio
\re{OR}:

\begin{proposition}
Let $\xi=\xi(\bpsi)$ be a measure \re{xi} of joint effects,
marginal effects or interaction among the risk factors in $J$. Assume
that it is
a function of the three\vadjust{\eject} quantities $a$, $b$ and $c$ in \re{abc}. The
derivative of $\xi$ with respect to the parameter vector $\bpsi$, is
then given by
\[
\bD= \frac{d\xi(\bpsi)}{d\bpsi} = \frac{\partial\xi}{\partial a}\bA+ \frac{\partial\xi}{\partial b}\bB+
\frac{\partial\xi}{\partial c}\bC,
\]
where
\begin{align*}
\bA&= \frac{d a}{d\scbfm{\psi}} = \OR_{(\scbfm{1},\sbv_K)}\bone_{\le
(\scbfm{1},\sbv_K)},\\
\bB&= \frac{d b}{d\scbfm{\psi}} = \sum_{\scbfm{0}\le\scbfm{u}\le\scbfm
{1}\atop|\scbfm{u}|\le i-1}
\sum_{\scbfm{0}\le\sbw\le\sbu} (-1)^{|\sbu|-|\sbw|}\OR_{(\sbw,\sbv
_K)}\bone_{\le(\sbw,\sbv_K)}\\
&= \sum_{\scbfm{0}\le\scbfm{w}\le\scbfm{1}\atop|\scbfm{w}|\le i-1} \OR
_{(\sbw,\sbv_K)}\bone_{\le(\sbw,\sbv_K)} (-1)^{i-1-|\sbw|}
{|J|-1-|\sbw|\choose i-1-|\sbw|}, \\
\bC&= \frac{d c}{d\scbfm{\psi}} = \OR_{(\sbfm{0},\sbv_K)}\bone_{\le
(\sbfm{0},\sbv_K)}.
\end{align*}
\end{proposition}

With a slight abuse of notation, we included binary vectors $\bu$ and
$\bw$ of length $|J|$ (not $p$) in the definition of $\bB$.
In the last step of this equation we interchanged order of summation
between $\bu$ and $\bw$ and then simplified the expression for the
inner sum, similarly as in Appendix \ref{appb}.

\end{appendix}

% structpyb loaded by giedrius.virsilas, 2017-04-21 11:27:34

%
\end{document}